\documentclass[final,5p,times,twocolumn]{elsarticle}
\usepackage{amsmath}
\usepackage{amsfonts}
\usepackage{amssymb}
\usepackage{graphicx}
\usepackage{dcolumn}
\usepackage{color}
\journal{Arxiv}

\begin{document}

\begin{frontmatter}

\title{Evidence of a two--dimensional grain sliding regime in the
plastic deformation of micron and sub--micron films.}

\author[utal]{Miguel Lagos}
\ead{mlagos@utalca.cl}
\author[utal]{V\'ictor Conte}
\ead{victor.conte@gmail.com}
\author[Simap,UCh]{Michel Ignat}
\ead{mignat@ing.uchile.cl}
\address[utal]{Facultad de Ingenier\'\i a, Universidad de Talca,
Campus Los Niches, Camino a los Niches Km 1, Curic\'o, Chile}
\address[Simap]{on leave from, Simap CNRS,Grenoble-INP,BP 75,
38402 Saint Martin d'H\`eres, France}
\address[UCh] {DFI, DIMEC, Facultad de Ciencias F\'isicas y
Matem\'aticas, Universidad de Chile, Beauchef 850, Santiago, Chile}

\begin{abstract}
Copper and aluminium sub--micron films are structured as columnar
arrays of grains traversing the entire film, and hence are virtual
two--dimensional polycrystalline solids. A closed--form solution for
the two--dimensional plastic flow is obtained assuming that grain
boundary sliding is the dominating mechanism for deformation. The
theoretical results obtained here are in close agreement with
experimental data appeared in the literature, which suggests further
work on the subject.
\end{abstract}


\end{frontmatter}

Tensile and compressive stresses of thermal origin can reach hundreds
of MPa in the small--scale structures of a complex microelectronic
device, either on service or in the fabrication process. Besides, the
strong functional interdependence of the numerous constituents of the
system leaves no tolerance for local strain failures. The structural
dimensions of current microsystems are in the length scale of 100 nm
or less. Hence, reliable design ensuring no risk of stress--induced
catastrophic defect demands precise knowledge of the mechanical
properties of sub--micron metallic films, together with a good
understanding of the response of the film to strongly varying
stresses. This has opened a new field of research in materials science
because the mechanical behaviour of thin films has proven to be very
different from that of their bulk counterparts.

Normal--view and cross--sectional micrography has shown that
polycrystalline sub--micron films have grain sizes larger or of the
order of the film thickness, with grains traversing the entire film
\cite{XiangVlassak,Gruber}. Typical films are structured as columnar
arrays of grains which in the plane of the film exhibit random
equiaxed shapes and crystallographic orientations. However, partial
crystallographic ordering is detected in the direction normal to the
film plane. The grains of copper sub--micron films, either
self--standing or on a supporting substrate, show strong tendency to
orient the (111) crystal direction along the normal to the film
surfaces \cite{XiangVlassak,Gruber}.

A sub--micron metallic film of this kind configures a physical
realization of a two--dimensional polycrystalline solid. The interest
of this goes beyond the technical applications because provides means
for the laboratory testing of theoretical models for plastic flow in a
reduced dimensionality.  Techniques for the fabrication and mechanical
testing of self standing beams of sub--micron thickness have been
developed in recent years \cite{Malhaire}. Besides this, a neat
solution for the problem of drawing out the intrinsic film mechanical
properties when testing thin films on substrates has appeared quite
recently in the literature. In it the true stresses exerted in both
principal directions of a metallic film, deposited on a compliant
polymer substrate, are accurately measured by diffracting {\it in
situ} the X--rays coming from a synchrotron source \cite{Gruber}. The
diffraction pattern reveals the distortions of the crystal cells of
the metallic film, which are proportional to the true stresses
operating on the film. The stabilising effect of the substrate permits
relatively large plastic tensile and compressive deformation of films
as thin as 80 nm without the risk of buckling. Because of the mismatch
of the Poisson ratios of the metallic film and the substrate, any
strain $\varepsilon$ in the longitudinal direction is associated not
only to a stress $\sigma_L$ in that direction, but also to a stress
$\sigma_T$ in the transversal one. This is not a serious shortcoming
because the method gives both stresses, $\sigma_L$ and $\sigma_T$, and
what matters is the difference $\sigma_L-\sigma_T$ between them, as
will become clear in what follows.

Experiments on copper \cite{Gruber} and aluminium \cite{Malhaire}
micron and sub--micron films show that the plastic deformation starts
with a strong but brief hardening period, going to not beyond 0.5\%
strain, followed by a plateau or a slowly softening regime extending
up to failure. Simulations by von Blanckenhagen et
al.~\cite{Blackenhagen} of the response to an applied stress of the
dislocation sources in a representative columnar grain, and the
subsequent dislocation dynamics, explains very well the initial
strengthening plastic deformation, fitting both the measured
stress--strain curves and dislocation densities. However, for strains
exeeding 0.5\% the simulation gives too large stresses and does not
explain the plateau, suggesting a crossover with another deformation
mechanism. We provide here supporting evidence that the latter higher
strain mechanism is two--dimensional grain sliding.

If we assume that grain boundary sliding is the dominant mechanism for
the plastic deformation of the film, we can reduce the general
theoretical formalism of Ref.~\cite{LagosConte} to two--dimensions for
analysing the data. The basic hypothesis of the model is that adjacent
grains can slide past each other over long distances by effect of the
shear stresses actuating in their shared boundaries, accommodating at
the same time their shapes by internal mechanisms to prevent voids at
the interfaces and preserve matter continuity. The sliding--induced
stress fields associated to grain shape accommodation are of little
relevance because are assumed much weaker than the shear stress
causing grain sliding. The latter is linearly related with the
relative speed $|\Delta\vec v|$ between the adjacent grains and has a
threshold $\tau_c$, below which no elementary sliding process can
occur \cite{Qi}.

\begin{figure}[h!]
\begin{center}
\includegraphics[width=5.5cm]{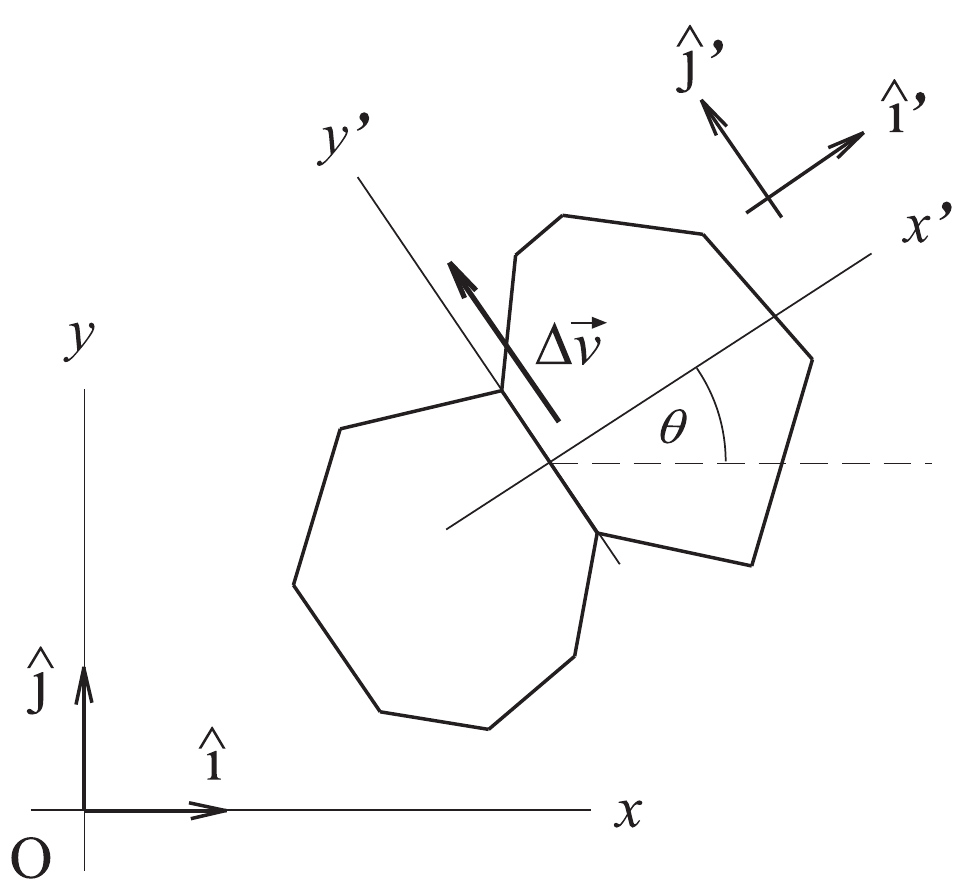}
\caption{\label{Fig1} Schematic representation of two adjacent grains
showing the local $(x'y')$ and main $(xy)$ frames of reference, the
corresponding unitary vectors and the relative velocity
$\Delta\vec v$.}
\end{center}
\end{figure}

Fig.~\ref{Fig1} represents a pair of adjacent two--dimensional
grains. The $x'$ and $y'$ axes of the local frame of reference
$(x'y')$, associated to the unitary vectors $\hat{\text\i}'$ and
$\hat{\text\j}'$, are normal and parallel to the grain boundary shared
by the grains, respectively. The main frame of reference $(xy)$, with
unitary vectors $\hat{\text\i}$ and $\hat{\text\j}$, has its axes in
the principal directions of the stress tensor. Denoting
$\sigma_{i'j'}$, $i',j'=x',y'$, the components of the stress tensor in
the $(x'y')$ frame of reference, the relative velocity $\Delta\vec v$
of the two grains is

\begin{equation}
\begin{aligned}
\Delta\vec v =
\begin{cases}
\mathcal{Q}\, (\sigma_{x'y'}-s'\tau_c)\,\hat{\text\j}'\quad
&\text{if}\quad \sigma_{x'y'}>\tau_c \\
0, &\text{ otherwise},  
\end{cases}
\label{E1}
\end{aligned}
\end{equation}

\noindent
where $\mathcal{Q}$ is a proportionality coefficient and
$s'=\sigma_{x'y'}/|\sigma_{x'y'}|$ is the sign of the shear stress
$\sigma_{x'y'}$. The term $s'\tau_c$ ensures that $\Delta\vec v=0$
when $|\sigma_{x'y'}|=\tau_c$. This expresion for $\Delta\vec v$ has
proven to hold with great accuracy for several aluminium, titanium and
magnesium alloys \cite{LagosRetamal1}. The coefficient $\mathcal{Q}$
must not depend on either the shear stresses or the orientation of the
grain boundary, therefore its dependence on the normal stresses is
only via the hydrostatic pressure invariant
$p=-(\sigma_{x'x'}+\sigma_{y'y'})/2$
\cite{Lagos1,Lagos2}.

Replacing $\sigma_{x'y'}=-(\sigma_x-\sigma_y)\sin\theta\cos\theta$,
where $\sigma_x$ and $\sigma_y$ are the principal stresses and
$\theta$ is the rotation angle defined in Fig.~\ref{Fig1}, and
$\hat{\text\j}'=-\sin\theta\,\hat{\text\i}
+\cos\theta\,\hat{\text\j}$, it gives

\begin{equation}
\Delta\vec v=\mathcal{Q}[-(\sigma_x-\sigma_y)\sin\theta\cos\theta
+s'\tau_c](-\sin\theta\,\hat{\text\i} +\cos\theta\,\hat{\text\j}),
\label{E2}
\end{equation}

\noindent
which presumes that

\begin{equation}
\sin (2\theta)\ge \frac{2\tau_c}{|\sigma_x-\sigma_y|}
\equiv\sin (2\theta_c),
\label{E3}
\end{equation}

\noindent
and otherwise $\Delta\vec v=0$. A grain boundary whose normal
$\hat{\text\i}'$ subtends with the principal direction $x$ or $y$ an
angle smaller than $\theta_c$ is not able to slide because the
in--plane shear stress is below $\tau_c$, no matter how strong the
external forces may be.

\begin{figure}[h!]
\begin{center}
\includegraphics[width=8.5cm]{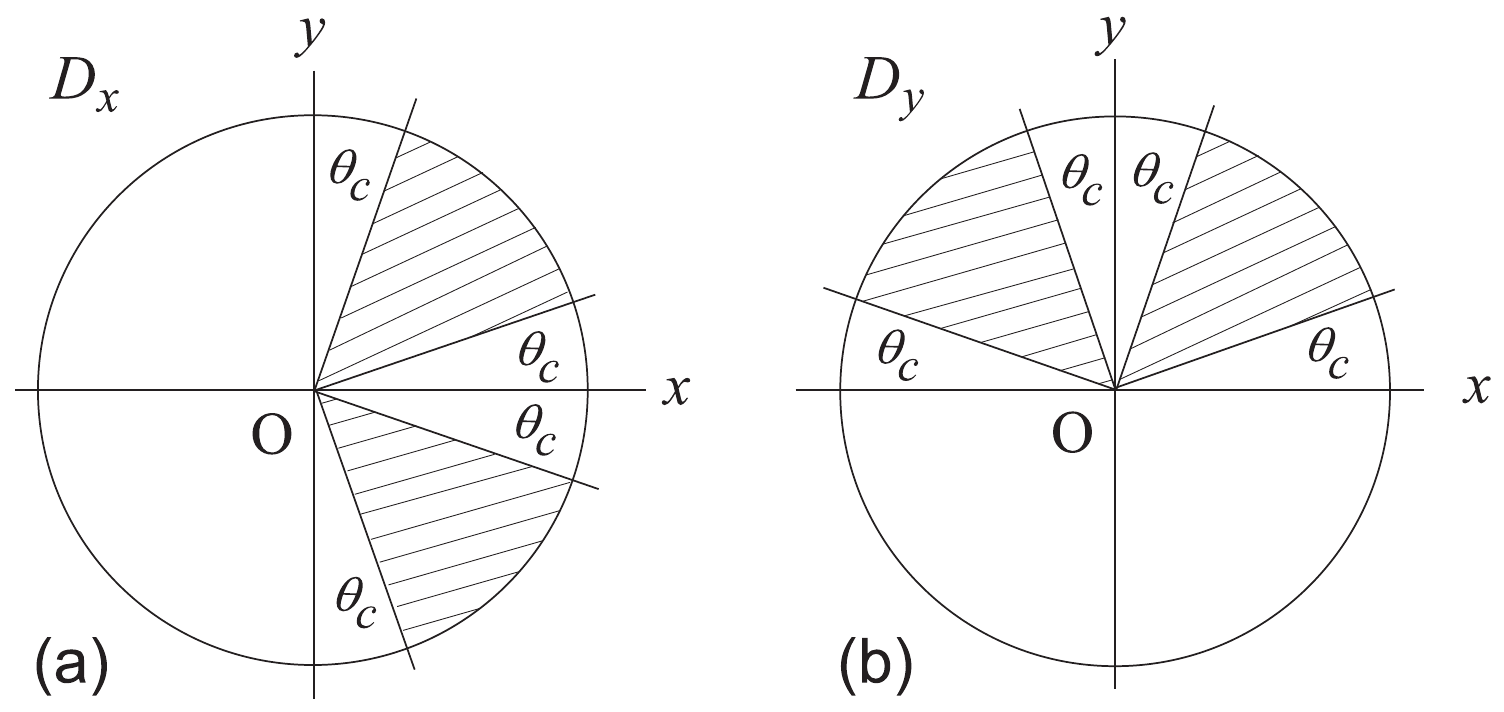}
\caption{\label{Fig2} The shaded areas represent the integration
domains $D_x$ (a) and $D_y$ (b) for the variable $\theta$ in
integrals (\ref{E6}).}
\end{center}
\end{figure}

To link these equations with the velocity field $\vec v(x,y)$ of the
plastically flowing film, consider two points at $(x,y)$ and
$(x+\delta x,y)$. The macroscopically small segment $\delta x$
intersects a large number $n$ of grain boundaries, and then $\delta
x=nd$, where $d$ is the mean grain size. The relative velocity between
the starting and final points of the segment $\delta x$ is the sum of
the $n$ relative velocities between the consecutive grains it passes
through. Thus

\begin{equation}
\frac{\vec v(x+\delta x,y)-\vec v(x,y)}{\delta x}
=\frac{1}{nd}\sum_{k=1}^n \Delta\vec v (k),
\label{E4}
\end{equation}

\noindent
where $k$ numbers the successive grain boundaries intersecting $\delta
x$. But for the factor $d$ in the denominator, the right--hand--side
of this equation defines an average. Consequently, in the proper
limit,

\begin{equation}
\frac{\partial\vec v}{\partial x_i}
=\frac{1}{d}\langle\Delta\vec v\rangle_i,
\quad  i=x,y\quad\text{or}\quad x_i=x,y\, ,
\label{E5}
\end{equation}

\noindent
where symbol $\langle\dots\rangle_i$ means the average over all
boundary orientations $\theta$ compatible with $x_i\ge 0$. With the
latter restriction and recalling condition (\ref{E3}) this means

\begin{equation}
\frac{\partial\vec v}{\partial x_i}
=\frac{1}{\pi d}\int_{D_i}\, d\theta\,\Delta\vec v,
\quad i=x,y\quad\text{or}\quad x_i=x,y\, ,
\label{E6}
\end{equation}

\noindent
where the integration domains $D_x$ and $D_y$ are those shown in
Fig.~\ref{Fig2}. Solving one obtains that

\begin{equation}
\begin{aligned}
&\frac{\partial v_x}{\partial x}=-\frac{\partial v_y}{\partial y}
=s\frac{2\mathcal{Q}\tau_c}{\pi d}\left(\frac{\cos^3\theta_c
-\sin^3\theta_c}{3\sin\theta_c\cos\theta_c}
-\cos\theta_c+\sin\theta_c\right),\\
&\frac{\partial v_x}{\partial y}=\frac{\partial v_y}{\partial x}=0.
\end{aligned}
\label{E7}
\end{equation}

\noindent
where

\begin{equation}
\theta_c=\frac{1}{2}
\arcsin\left(\frac{2\tau_c}{|\sigma_x-\sigma_y|}\right),
\quad s=\frac{\sigma_x-\sigma_y}{|\sigma_x-\sigma_y|}.
\label{E8}
\end{equation}

These equations assume that the mean grain size $d$ is conserved, and
yield a complete set of equations for the velocity field $\vec v$ of
the material medium in terms of the stresses. From them

\begin{equation}
\nabla\times\vec v=0,\quad\nabla\cdot\vec v=\frac{\dot A}{A}=0,
\label{E9}
\end{equation}

\noindent
which means that the plastic flow is laminar and conserves the film
specific area $A$. The strain rate tensor
$\dot\varepsilon_{ij}=(\partial v_i/\partial x_j +\partial
v_j/\partial x_i)/2$ follows directly from Eqs.~(\ref{E7}), explicitly

\begin{equation}
\begin{aligned}
&\dot\varepsilon_{xx}=-\dot\varepsilon_{yy}
=\frac{2\mathcal{Q}\tau_c}{\pi d}\left(\frac{\cos^3\theta_c
-\sin^3\theta_c}{3\sin\theta_c\cos\theta_c}
-\cos\theta_c+\sin\theta_c\right),\\
&\dot\varepsilon_{xy}=0, \quad (\sigma_x\ge\sigma_y)
\end{aligned}
\label{E10}
\end{equation}

\noindent
for the biaxial deformation with principal axes $x$ and $y$.

Eqs.~(\ref{E10}) and (\ref{E8}) show two very important properties,
specific to the two--dimensional plastic flow by grain sliding. They
are:\\ (i) The stresses depend only on the strain rates, and not on
the strains. Hence no strain hardening or softening of the material is
expected in a bidimensional grain sliding plastic regime.\\
(ii) The ratio between the true strain rates and the coefficient
$\mathcal{Q}$ depends on the principal stresses only through their
difference $\sigma_x-\sigma_y$. As explained before, the coefficient
$\mathcal{Q}$ is expected to depend on the pressure invariant
$p=-(\sigma_x+\sigma_y)/2$, however this dependence seems to be rather
weak at room temperature, as revealed by the experimental data shown
next.

\begin{figure}[h!]
\begin{center}
\includegraphics[width=8.5cm]{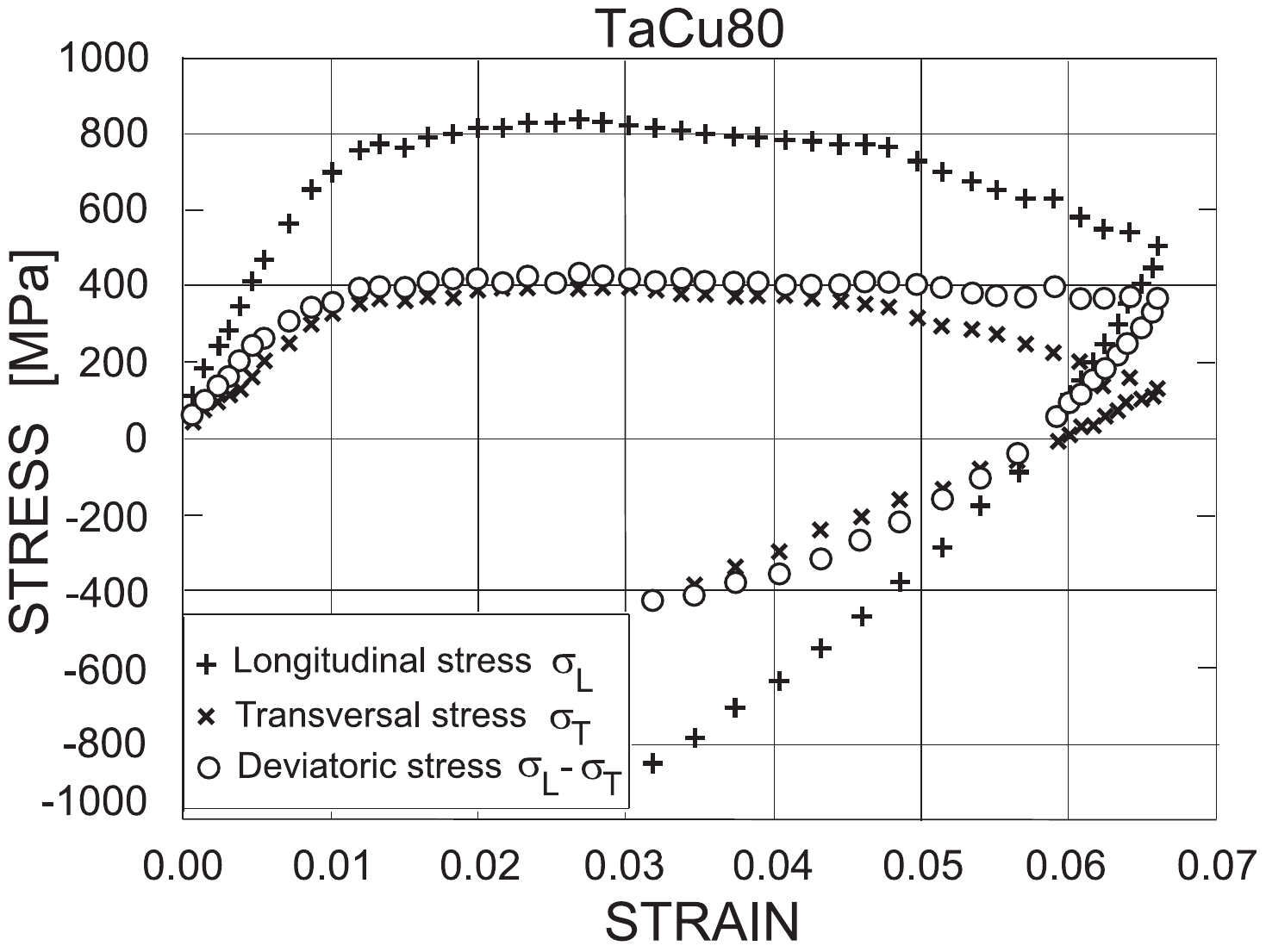}
\caption{\label{Fig3} Data of Gruber et al.~\cite{Gruber} for a copper
film of $67\,\text{nm}$ thick on a polyimide substrate, strained at
constant strain rate. The film has a passivating tantalum layer of 9
nm thick. Crosses represent the original data and circles the
difference $\sigma_L-\sigma_T$ between the two principal stresses.
Within the experimental uncertainty, $\sigma_L-\sigma_T$ in the
tensile region is constant for $\varepsilon >0.01$.}
\end{center}
\end{figure}     

Fig.~\ref{Fig3} shows the data of Gruber et al. \cite{Gruber} on a
copper film of $67\,\text{nm}$ thickness and mean grain size
$d=104\,\text{nm}$ (measured in the film plane), deposited on a
polyimide substrate and strained at constant strain rate. The film
incorporates a capping tantalum layer of 9 nm thick. The principal
stresses $\sigma_L$ and $\sigma_T$, which can be assimilated to
$\sigma_x$ and $\sigma_y$ in Eqs.~(\ref{E10}) and (\ref{E8}), were
measured by the synchrotron X--ray diffration method.  The graph shows
that both $\sigma_L$ and $\sigma_T$ vary with the strain
$\varepsilon$, which is assimilable to $\varepsilon_{xx}$ in the
previous equations. The longitudinal stress rapidly increases to a
maximum $\sigma_L=846\,\text{MPa}$ at $\varepsilon =0.027$ and then
decreases monotonically to $\sigma_L=515\,\text{MPa}$ when the tensile
test ends at $\varepsilon =0.066$. The transversal stress $\sigma_T$
displays a similar behaviour, with maximum $\sigma_T=404\,\text{MPa}$
at $\varepsilon =0.027$ and $\sigma_T=140\,\text{MPa}$ at $\varepsilon
=0.066$. Both stresses vary a few hundreds MPa and are far from
constant.

Instead, the difference $\sigma_L-\sigma_T$ (white circles) increases
rapidly in the interval $0\le\varepsilon <0.015$ and then remains
nearly constant up to the end of the tensile test at
$\varepsilon =0.066$. Recalling Eqs.~(\ref{E10}) and (\ref{E8}), this
evidences that the tensile test was carried out at a constant strain
rate. To be precise, the cross--head displacement of the mechanical
testing machine was increased in small steps because the registration
of a diffraction picture takes one or two minutes. Also, the flat
region of the curve $\sigma_L-\sigma_T$ vs.~$\varepsilon$ has a slight
negative slope, which can be explained because what was kept constant
in the experiment was the engineering strain rate, while
$\dot\varepsilon_{xx}$ in Eq.~(\ref{E10}) is the true strain
rate. When the engineering strain rate is maintained constant, the
true strain rate diminishes with strain, and the same does
$\sigma_x-\sigma_y$, as given by Eq.~(\ref{E10}). It is usually
argued that plastic deformation is driven by the deviatoric principal
stresses $\sigma_i+p$, where the hydrostatic pressure $p$ is the
negative average of the normal stresses. In two dimensions the
deviatoric stress is $(\sigma_x-\sigma_y)/2$. In the insets of the
figures we disregarded the factor 1/2 and called the deviatoric stress
to $\sigma_L-\sigma_T$.

Upon unloading, Fig.~\ref{Fig3} shows that $\sigma_L-\sigma_T$ varies
linearly with $\varepsilon$, with a slope close to the one displayed
by the elastic part of the loading curve. On compression,
$\sigma_L-\sigma_T$ converges rather slowly to a compressive net
stress $419\,\text{MPa}$, which is a quantity amazingly close to the
tensile strain $\sigma_L-\sigma_T$ at the same strain in the tensile
plateau of the experimental curve. This shows that the material
undergoes no final strain strengthening or weakening, that is, no
Bauschinger effect seems to occur.
   
\begin{figure}[h!]
\begin{center}
\includegraphics[width=8.5cm]{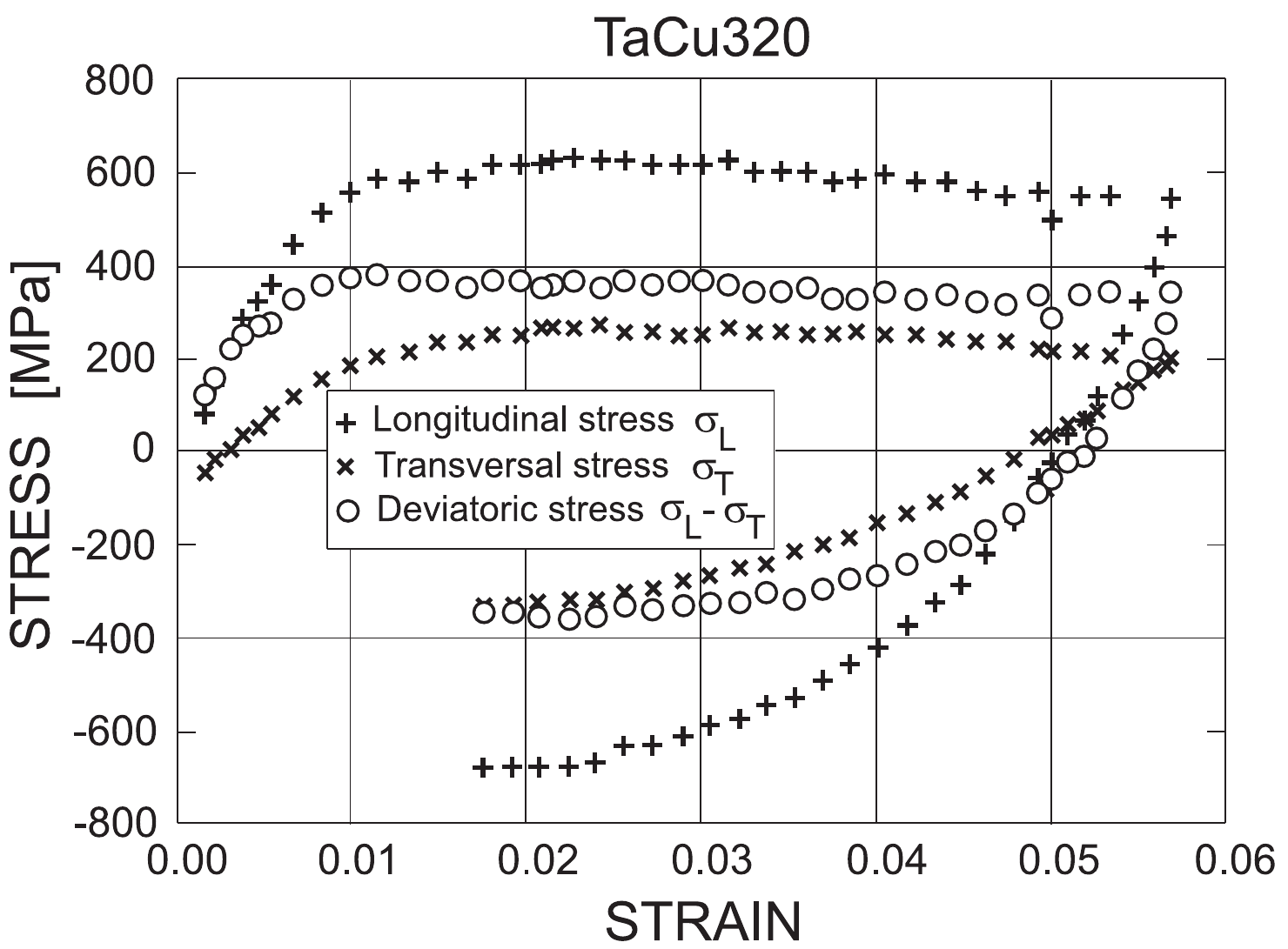}
\caption{\label{Fig4} Data of Gruber et al.~\cite{Gruber} for a
$251\,\text{nm}$ thick copper film and $d=297\,\text{nm}$ on a
polyimide substrate, strained at constant strain rate. The film has a
passivating tantalum layer of 8 nm thick. The difference
$\sigma_L-\sigma_T$ between the two principal stresses keeps almost
constant in the tensile region for $\varepsilon >0.01$. Upon
compression, $\sigma_L-\sigma_T$ go to a constant of the same
magnitude.}
\end{center}
\end{figure}

\begin{figure}[h!]
\begin{center}
\includegraphics[width=8.5cm]{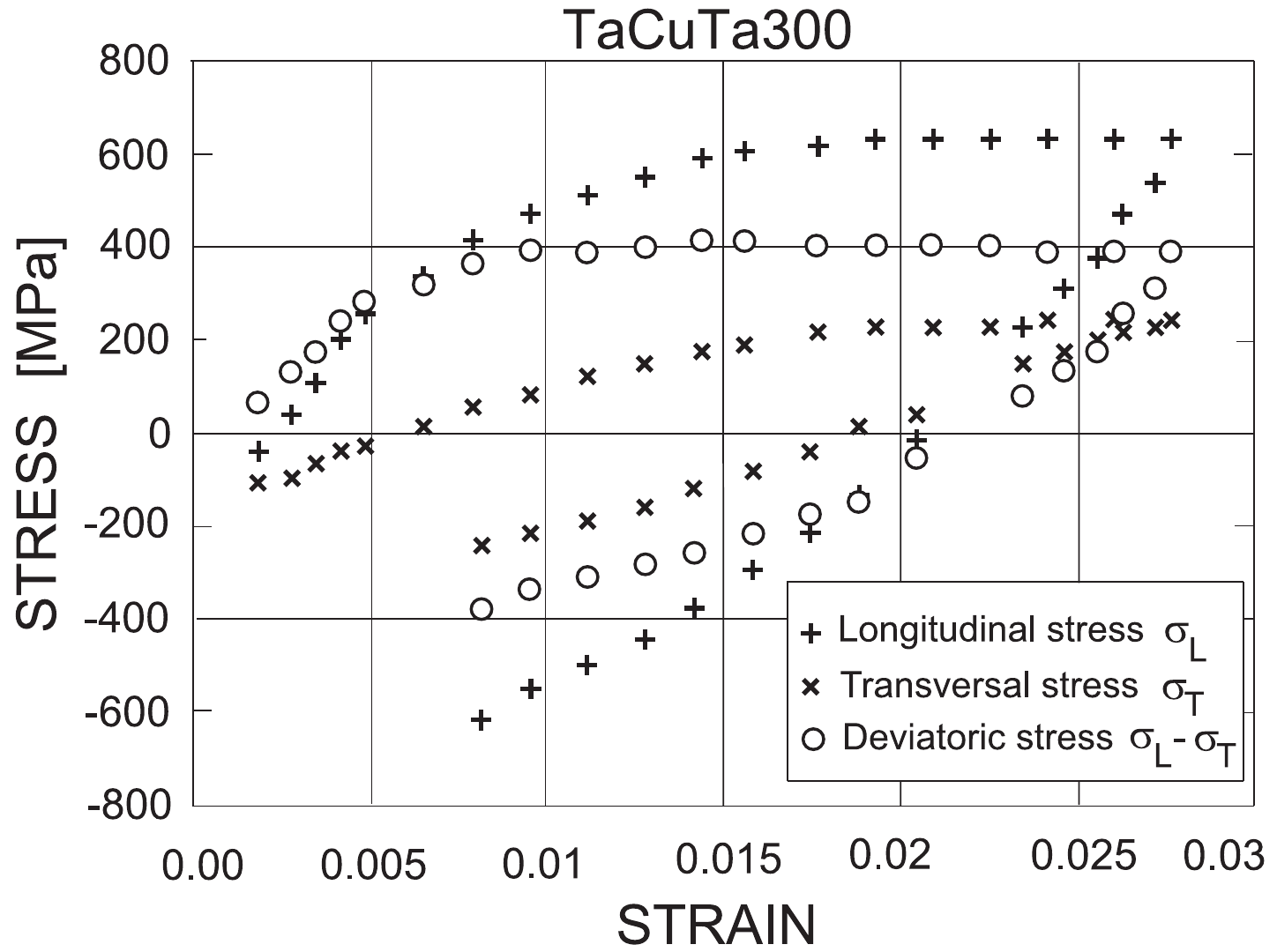}
\caption{\label{Fig5} Data of Gruber et al.~\cite{Gruber} on a
copper film of $255\,\text{nm}$ thick  on a polyimide substrate,
strained at constant strain rate. The film has passivating tantalum
layers of 8 nm thick in both sides. The difference $\sigma_L-\sigma_T$
between the two principal stresses exhibits the same trends of
Figs.~\ref{Fig3} and \ref{Fig4}.}
\end{center}
\end{figure}

\begin{figure}[h!]
\begin{center}
\includegraphics[width=8.5cm]{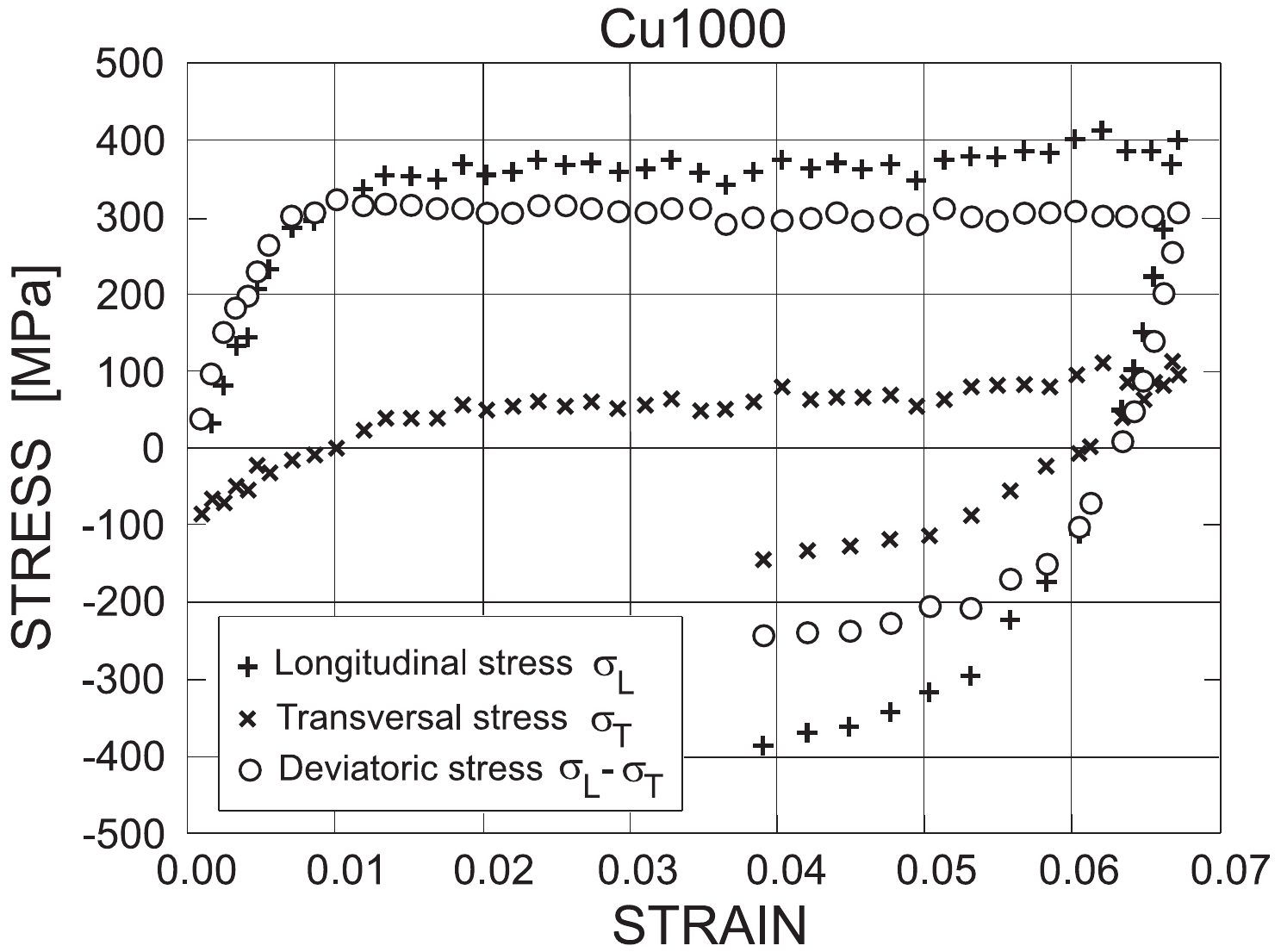}
\caption{\label{Fig6} Data of Gruber et al.~\cite{Gruber} on a copper
film of $765\,\text{nm}$ thick and $d=423\,\text{nm}$ on a polyimide
substrate, strained at constant strain rate. The film has no
passivating layers. The difference $\sigma_L-\sigma_T$ of the principal
stresses exhibits trends similar to those of
Figs.~\ref{Fig3}--\ref{Fig5}.}
\end{center}
\end{figure}

Figs.~\ref{Fig4} and \ref{Fig5} displays the data given by similar
experiments with copper films of different thicknesses and mean grain
sizes, and with one or two passivating tantalum layers of 8 nm thick.
As in Fig.~\ref{Fig3}, and by the same reason, in both cases the
plateau exhibits a small negative slope. Upon compression,
$|\sigma_L-\sigma_T|$ converges rather slowly to near the same stress
of the tensile plateau. Again the material seems to undergo no
Bauschinger effect. Fig.~\ref{Fig6} reveals results of a similar
experiment, conducted now with a bare copper film of $765\,\text{nm}$
thick on polyimide, with no passivating coatings, strained at constant
strain rate. Again $\sigma_L-\sigma_T$ keeps constant for
$\varepsilon >0.01$ and shows a more regular behaviour than the
original data on $\sigma_L$ and $\sigma_T$.

\begin{figure}[h!]
\begin{center}
\includegraphics[width=8.5cm]{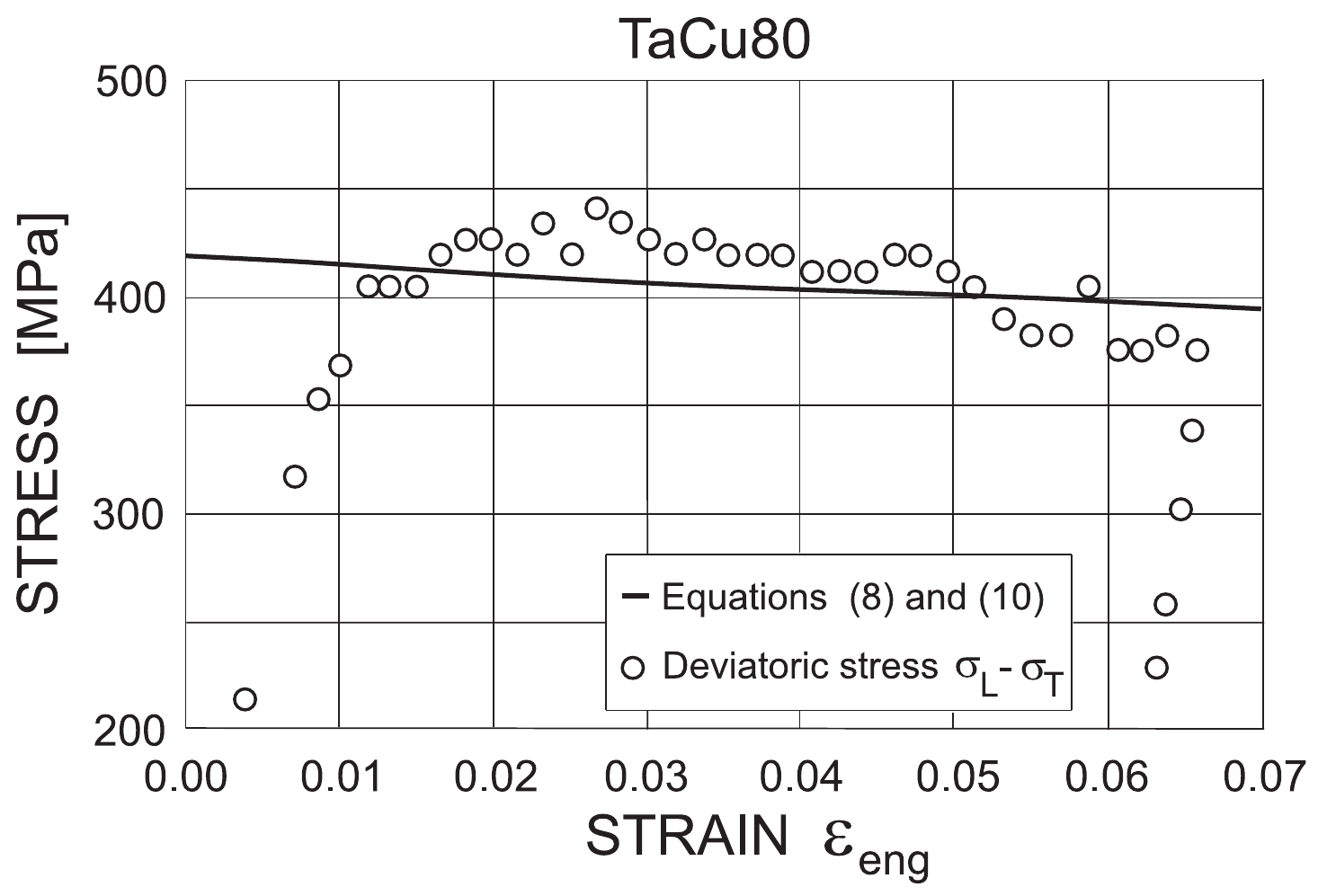}
\caption{\label{Fig7} Expansion of the tensile branch of
Fig.~\ref{Fig3}. Circles represent $\sigma_\text{L}-\sigma_\text{T}$,
with $\sigma_\text{L}$ and $\sigma_\text{T}$ taken from experiment.
The solid line is given by Eqs.~(\ref{E10}) and (\ref{E8}) with $\sigma_x-\sigma_y=\sigma_\text{L}-\sigma_\text{T}$ and under the
assumption that the engineering strain rate
$\dot\varepsilon_{\text{eng}}$ was kept constant all over the tensile
run.}
\end{center}
\end{figure}

Fig.~\ref{Fig7} displays and expanded view of the tensile branch
appearing in Fig.~\ref{Fig3}. Together with the experimental data, the
figure shows the curve of true stress vs.~engineering strain given by
Eqs.~(\ref{E10}) and (\ref{E8}) under the assumption that the
engineering strain rate was kept constant in the tensile run. If the
engineering strain rate $\dot\varepsilon_{\text{eng}}$ is constant,
the true strain rate $\dot\varepsilon$ decreases with
$\varepsilon_{\text{eng}}$, and the same should happen to the true
stress $\sigma$. The curve was obtained from
$\varepsilon_{\text{eng}}=\dot\varepsilon_{\text{eng}}/\dot\varepsilon
-1$, where $\dot\varepsilon=\dot\varepsilon_{xx}$ is the function of
$\sigma_x-\sigma_y=\sigma_\text{L}-\sigma_\text{T}$ given by
Eqs.~(\ref{E10}) and (\ref{E8}). Just to show an example of the
procedure we adopted $\tau_c=15\,\text{MPa}$ and
$2\mathcal{Q}\tau_c/(\pi d\dot\varepsilon_{\text{eng}})=0.12$.

The two--dimensional plastic behaviour of submicron films has a
special interest from a theoretical viewpoint. Making the same
hypotheses and reasoning the same way as done to obtain
Eq.~(\ref{E10}), but now in a medium of three--dimensional polyhedra,
one finds out that $\nabla\cdot\vec v\ne 0$ \cite{LagosConte}. Hence
the three--dimensional model yields non conservation of the specific
volume. As volume variations can only be supported by the elasticity
of the grains, one can recall Hooke's law and write $\nabla\cdot\vec
v=-(1/B)\dot p$, where $B$ is the bulk elastic modulus and $\dot p$
the pressure variation rate. This new equation determines that the
stresses are not independent dynamical variables, plastic deformation
is essentially a time--dependent problem and the stresses turn out
dependent of the strain components \cite{LagosConte}. Hence the
three--dimensional version of the model yields strain strengthening or
softening.  The two--dimensional case is free of these complications,
and thus sub--micron films are particularly suited for testing
experimentally the hypotheses of the model for plastic deformation by
grain sliding, which should hold the same in two-- or
three--dimensions.


\begin{thebibliography}{99}

\bibitem{XiangVlassak} Y. Xiang, J.J. Vlassak, Acta Mater. 54 (2006)
5449-–5460.

\bibitem{Gruber} P. A. Gruber, J. B\"ohma, F. Onuseit, A. Wanner,
R. Spolenak, E. Arzt, Acta Mater. 56 (2008) 2318-–2335. 

\bibitem{Malhaire} C. Malhaire, C. Seguineau, M. Ignat, C. Josserond,
L. Debove, S. Brida, J.M. Desmarres, X. Lafontan, Rev. Sci. Instrum.
82, 023901 (2009).

\bibitem{Blackenhagen} B. von Blanckenhagen, E. Arzt, P. Gumbsch, Acta
Mater. 52 (2004) 773–784.

\bibitem{LagosConte} M. Lagos, V. Conte, Scripta Mater. 65 (2011)
1053-–1056.

\bibitem{Qi} Y. Qi, P.E. Krajewski, Acta Mater. 55 (2007)
1555.

\bibitem{LagosRetamal1} M. Lagos, C. Retamal, Phys. Scr. 81 (2010)
055601.

\bibitem{Lagos1} M. Lagos, Phys. Rev. Lett. 85 (2000) 2332.

\bibitem{Lagos2} M. Lagos, Phys. Rev. B 71 (2005) 224117.

\end{thebibliography}
\end{document}